\documentclass[aps,prl,twocolumn,showpacs,floatfix]{revtex4}
\usepackage{amssymb,amsmath,epsfig,graphicx}
\begin{document}

\author{Francisco J. Cao$^{\text{1,2}}$, Luis Dinis$^{\text{1}}$ and Juan M.R. Parrondo$^{\text{1}}$}

\affiliation{$^{\text{1}}$ Grupo Interdisciplinar de Sistemas Complejos (GISC) and Departamento de F\'{\i}sica At\'omica, Nuclear y
    Molecular.
    Universidad Complutense de Madrid.    E-28040 Madrid, Spain.\\
    $^{\text{2}}$ Observatoire de Paris, LERMA.
    61, Avenue de l'Observatoire, 75014 Paris, France.
    Laboratoire Associ\'e au CNRS UMR 8112.}

\title{\bf Feedback control in a collective flashing ratchet}

\begin{abstract}
An ensemble of Brownian particles in a feedback controlled
flashing ratchet is studied. The ratchet potential is switched on
and off depending on the position of the particles, with the aim
of maximizing the current. We study in detail a protocol which
maximizes the instant velocity of the center of mass of the
ensemble at any time. This protocol is optimal for one particle
and performs better than any periodic  flashing for ensembles of
moderate size, whereas is defeated by a random or periodic
switching for large ensembles.
\end{abstract}
\pacs{05.40.-a, 02.30.Yy}

\maketitle

Rectification of thermal fluctuations is becoming a major research
topic in non equilibrium statistical mechanics, with potential
applications in biology, condensed matter, and nanotechnology
\cite{reiman,linke}.

\begin{figure}[ht]
\[
\includegraphics[height=3cm]{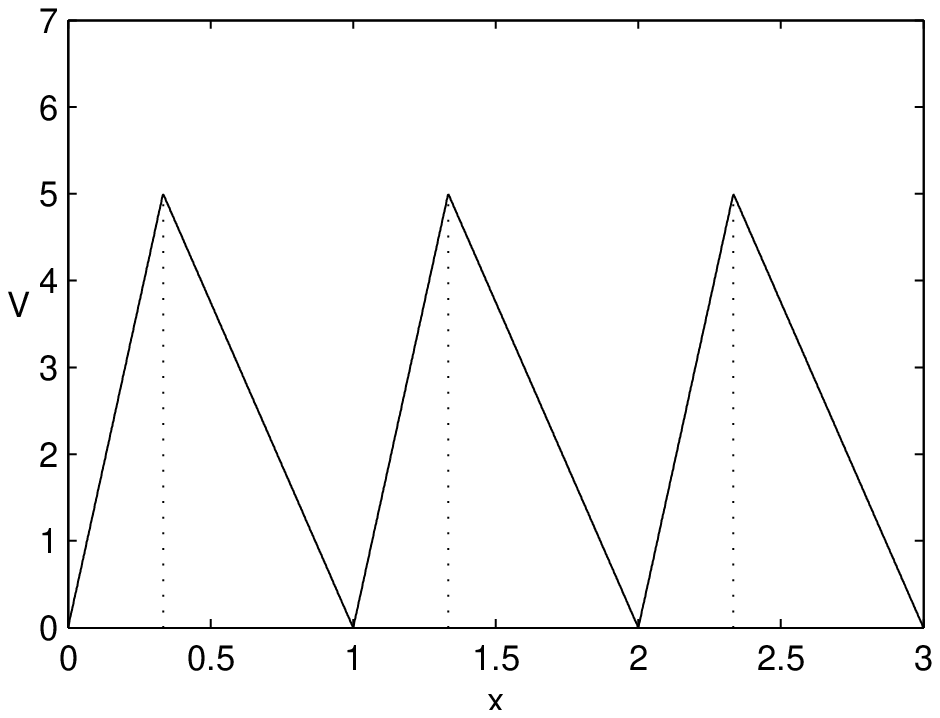}
\hspace{0.3cm}
\includegraphics[height=3cm]{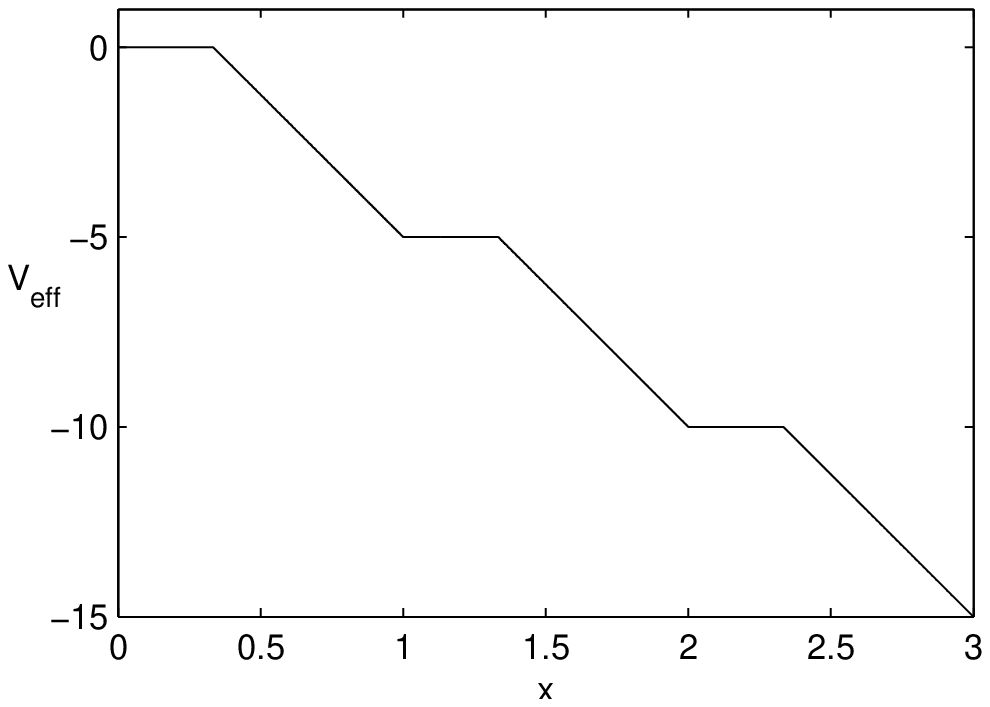}
\]
\caption{The ratchet potential ({\em left}) and the corresponding
effective potential for the one-particle controlled ratchet ({\em
right}), both for $V_0=5kT $ and $a=1/3$. (Units $L=1$, $D=1$ and
$kT=1$.)}
  \label{figpotentials}
\end{figure}

Most of these  rectifiers or Brownian ratchets work by introducing
an external time-dependent perturbation in an asymmetric
equilibrium system. In the case of rocking ratchets, the
perturbation is an AC uniform field, whereas for the flashing
ratchet \cite{ajdari,astumian} the perturbation consists of
switching on and off an asymmetric sawtooth potential, such as the
one depicted in Fig.~\ref{figpotentials} (left).

The models and applications studied so far have been focused on
periodic or random time-dependent perturbations \cite{reiman}. On
the other hand, in this Letter we study a {\em feedback controlled
perturbation}, i.e., an external force depending on the state of
the system.
Introducing control in Brownian ratchets is relevant for the
aforementioned applications.
Feedback control could be implemented in systems where particles are
monitored, as occurs in some experimental setups with colloidal particles
\cite{prostnature}.
Control theory is also of extreme relevance in biology \cite{mislum,kitano},
and most protein motors probably operate as control systems (see
\cite{serwer} for a specific example).
Finally, controlled ratchets are relevant from a theoretical
point of view. The idea of rectifying thermal noise was originally
introduced by Smoluchowski \cite{smol} and Feynmann \cite{feyn} in
relation to the Maxwell demon. Recently, Touchette and Lloyd
have pointed out that the original Maxwell demon can be considered
a feedback control system and found that thermodynamics imposes
some limitations to control \cite{lloyd}.

Using controlled Brownian ratchets one can build models to check
these limitations and understand how information can be used to
increase the performance of a system. For instance, it is not hard
to prove that some single-particle ratchets become Maxwell demons
if information on the position of the Brownian particle is
available \cite{b1}. In this Letter we will study the more
involved problem of a collective flashing ratchet, focusing on the
induced current of particles.

Consider a flashing ratchet \cite{ajdari,astumian} consisting of a
Brownian particle in an asymmetric sawtooth potential, such as the
one depicted in Fig.~\ref{figpotentials} (left). If the potential
is flashed in a periodic or random way, the particle exhibits a
systematic motion to the right
\cite{reiman,ajdari,astumian}. However, if we know where the
particle is at any time, there is  a better switching strategy. We
could switch on the potential whenever the force is positive and
switch it off if the force is negative. Under this switching
protocol, the particle feels the effective non periodic potential
depicted in Fig.~\ref{figpotentials} (right). Obviously, the
particle moves to the right and this is the optimal switching
policy, inducing larger currents than any periodic or random
switching.

The problem of control becomes less trivial for ratchets
consisting of many particles. Control strategies can induce an
effective coupling among particles and the system becomes a
coupled Brownian motor \cite{coupled}.

Consider an ensemble of $N$ overdamped Brownian particles at
temperature $T$ in an external asymmetric periodic potential $
V(x)$, that can be either on or off. The dynamics is
described by the Langevin equation:
\begin{equation}
\gamma \dot x_i(t) = \alpha(t) F(x_i(t)) + \xi_i(t) \;; \quad\quad
i = 1 \ldots N, \label{lang1}
\end{equation}
 where $ x_i (t)$ is the position of particle $ i $,
$\gamma$ is the friction coefficient and  $ \xi_i (t)$ are thermal noises
with zero mean and correlation $ \langle \xi_i(t)\xi_j(t') \rangle
= 2 \gamma k T \delta_{ij} \delta(t-t')$. The force is given by
$F(x) = -V'(x)$ where $V(x)$ is a periodic, $V(x+L) = V(x) $, asymmetric
potential defined by
\begin{equation}
 \label{Vondef} V(x) = \left\{
\begin{array}{lll}
\displaystyle \frac{V_0}{a}\frac{x}{L} & \mbox{if} & 0 \leq
\frac{x}{L}
\leq a  \\ && \\
 \displaystyle
 -\frac{V_0}{(1-a)}\left(\frac{x}{L}-a\right)+V_0 & \mbox{if} & a \leq \frac{x}{L} \leq 1 \\
 \end{array}
 \right.
\end{equation}
and depicted in Fig.~\ref{figpotentials}, left. Finally,
$\alpha(t)$ is a control parameter which we assume that can take
on the values 1 and 0, i.e., the only allowed operations on the
Brownian motor consist of switching on and off the potential
$V(x)$.

We will consider the following two switching strategies:
\begin{itemize}

\item {\em Periodic switching}: $ \alpha(t+\tau)= \alpha(t)
$, with $ \alpha(t)=1 $ for $ t \in [0,\tau/2)$, and $ \alpha(t)=0
$ for $ t \in [\tau/2,\tau)$. This case is equivalent to the
periodic flashing ratchet \cite{ajdari,astumian}, since particles
are independent.

\item \emph{Controlled switching}:
\begin{equation} \label{alphacont}
\alpha(t) = \Theta (f(t)) \quad \mbox{with} \quad
f(t) = \frac{1}{N}\sum_{i=1}^N F(x_i(t)),
\end{equation}
where $f(t)$ is the net force per particle and  $\Theta(y)$ is the
Heaviside function, $\Theta(y)=1$ if $y\geq 0$ and 0 otherwise. As
can be deduced from Eq.~(\ref{lang1}), this strategy maximizes the
instant velocity of the center of mass, $ \dot x_{cm}(t)=
\frac{1}{N} \sum_{i=1}^N \dot x_i(t)$. Particles are no longer
independent, due to the feedback control $\Theta (f(t))$.

\end{itemize}

\begin{figure}[ht]
\[
\includegraphics[height=4cm]{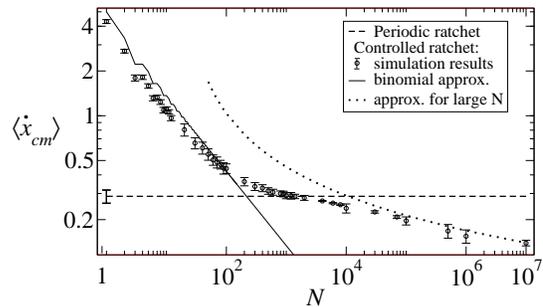}
\]
\caption{Average of the speed of the center of mass, $
\langle \dot x_{\rm cm} \rangle $, for simulations of the periodic
switching with optimal period $\tau_{\text{opt}}=0.05L^2/D$ ({\em
dashed line}); simulations of the controlled ratchet ({\em circles
with error bars}); the binomial approximation, Eq.~\eqref{bin},
({\em solid line}); and the large $ N $ approximation,
Eq.~\eqref{final}, ({\em dotted line}); all for $ V_0 = 5 k T $
and $ a = 1/3 $.  (Units $ L = 1 $, $ D = 1 $ and $ k T = 1 $.)
\label{J}}
\end{figure}

We first compare  numerical results for the two switching
strategies described above. In Fig.~\ref{J}, the center of mass velocity
is plotted as a function of $ N $
for the periodic switching with the optimal period
\cite{reiman,ajdari,astumian,nosefbr}, and for the controlled
switching, both for $ V_0 = 5 k T $ and $a = 1/3$. For this
potential, the optimal period has been found to be
$\tau_{\text{opt}}\simeq 0.05L^2/D$, giving $\langle \dot
x_{cm}\rangle\simeq 0.29D/L$. The controlled switching yields a
higher velocity than the periodic strategy only up to a certain
$N$ ($ N\simeq 1300$ for the previous values of the parameters).
We also see from simulations that the velocity goes to zero when $
N \to \infty $, in the controlled case. This is surprising at
first sight, because the controlled strategy given by
(\ref{alphacont}) maximizes the instant velocity $ \dot x_{\rm
cm}(t) $. However, this local maximization does not ensure good
results in the long term, as we have already shown for the
so-called paradoxical games and other deterministic systems
\cite{eplsr}.

The case $N=1$ can be solved analytically, since it consists of a
single particle moving in the effective potential depicted in the
rightmost plot of Fig.~\ref{figpotentials}, as mentioned above.
The corresponding stationary Fokker-Planck equation can be solved,
yielding
\begin{equation}
\langle \dot x(t) \rangle_{\rm st} =
  \frac{2DV_0(1-e^{\frac{-V_0}{kT}})/L}
  {2kT(1-a^2)(1-e^{\frac{-V_0}{kT}})+V_0a^2(1+e^{\frac{-V_0}{kT}})},
 \label{n1}
\end{equation}
where $ D = k T / \gamma $ is the diffusion coefficient. For $ V_0
= 5 k T $ and $ a =1/3$, this yields a stationary speed $\langle
\dot x\rangle_{st}\simeq 4.27 D/L$, more than $ 10 $ times larger
than the highest speed obtained with a periodic switching. For
$N=1$, in fact, this controlled switching is clearly optimal.

For arbitrary $N$, by summing and averaging the Langevin equations
(\ref{lang1}) with the prescription (\ref{alphacont}), one obtains
the following exact equation for the mean velocity of the center
of mass:
\begin{equation}
\left\langle \dot{x}_{\rm cm}(t) \right\rangle = \frac{1}{\gamma
}\left\langle \Theta (f(t)) \, f(t) \right\rangle. \label{jota}
\end{equation}
The net force per particle $f(t)$ can be written in terms of the
number $n(t)$ of particles in the interval $[0,aL]$. Notice also that
the system reaches a stationary regime, because there is not any explicit
dependence on time in the controlled switching case.

The average in the r.h.s. of Eq. \eqref{jota} can be
approximately computed in the stationary regime  making the following
two assumptions: {\em (i)} particles are statistically independent (i.e., the
stationary distribution has the form $ \rho_{st}(x_1, x_2,
\ldots, x_N) = \prod_{i=1}^N \rho_{st}(x_i) $)
and {\em (ii)} the probability that a given particle is in the
interval $[0,aL]$ is $a$. These are reasonable assumptions since
both are fulfilled by the two equilibrium distributions
corresponding to respectively switching on and off for a long
period of time. Under this approximation, the probability
distribution of the random variable $n(t)$ is a binomial
distribution in the stationary regime and the average velocity
becomes
\begin{equation}
 \left\langle \dot{x}_{\rm cm} \right\rangle_{\rm st} \simeq
\frac{V_0}{\gamma L N} \sum_{n<Na}^{N} \left( -\frac{n}{a}+
\frac{N-n}{1-a} \right)
\left( \begin{array}{c} N\\n\end{array} \right) a^n (1-a)^{N-n}.
\label{bin}
\end{equation}
This binomial approximation, Eq.~\eqref{bin}, has been tested for
various values of $ V_0 $ and $ a $ giving good results for small
$ N $ (see Fig.~\ref{J}). Eq.~\eqref{bin} is in fact exact for
$N=1$ in the limit $V_0/(kT)\to 0$ [as can be proven from
Eq.~\eqref{n1}]. This is expected since the equilibrium
distribution is then almost uniform both for the on and off
potentials. As a consequence, the binomial approximation gives
better results when decreasing $ V_0/(kT) $, not only for $ N = 1
$, but also for $ N > 1 $.

For sufficiently large values of $N$ ($ \gtrsim 10 $), the
binomial distribution can be approximated by a Gaussian, yielding
\begin{equation}
\left\langle \dot{x}_{\rm cm} \right\rangle_{\rm st} \simeq
\frac{1}{\sqrt{2\pi\sigma^2}}
\int_{-\infty}^{\infty} df\, \Theta(f)\, \frac{f}{\gamma}\,
e^{-\frac{f^2}{2\sigma^2}} = \frac{\sigma}{\gamma\sqrt{2\pi}},
\label{velbin}
\end{equation}
with $ \sigma^2  = V_0^2/[L^2 a
(1-a) N ] $.

One can study within this approximation the following general
mixed strategy depending on $f(t)$: the potential is switched on
with a probability $p_{\rm on}(f(t))$, and switched off with
probability $1-p_{\rm on}(f(t))$. Under the gaussian
approximation, the stationary velocity for this protocol is the
same as in Eq.~\eqref{velbin} but replacing $\Theta(f)$ by $p_{\rm
on}(f)$. Then it can be easily seen that the gaussian
approximation predicts that the optimal strategy is the one
considered in this paper, i.e., $p_{\rm on}(f)=\Theta (f)$.
However, this is not true for large $N$ (see Fig.~\ref{J}).

The binomial and the gaussian approximations fail in other respects for large
values of $N$. Both give $\left\langle
\dot{x}_{\rm cm} \right\rangle_{\rm st}\sim 1/\sqrt{N}$, whereas
numerical simulations show that $\left\langle \dot{x}_{\rm cm}
\right\rangle_{\rm st}$ decays more slowly.

\begin{figure} [ht]
\[
\includegraphics[height=3cm]{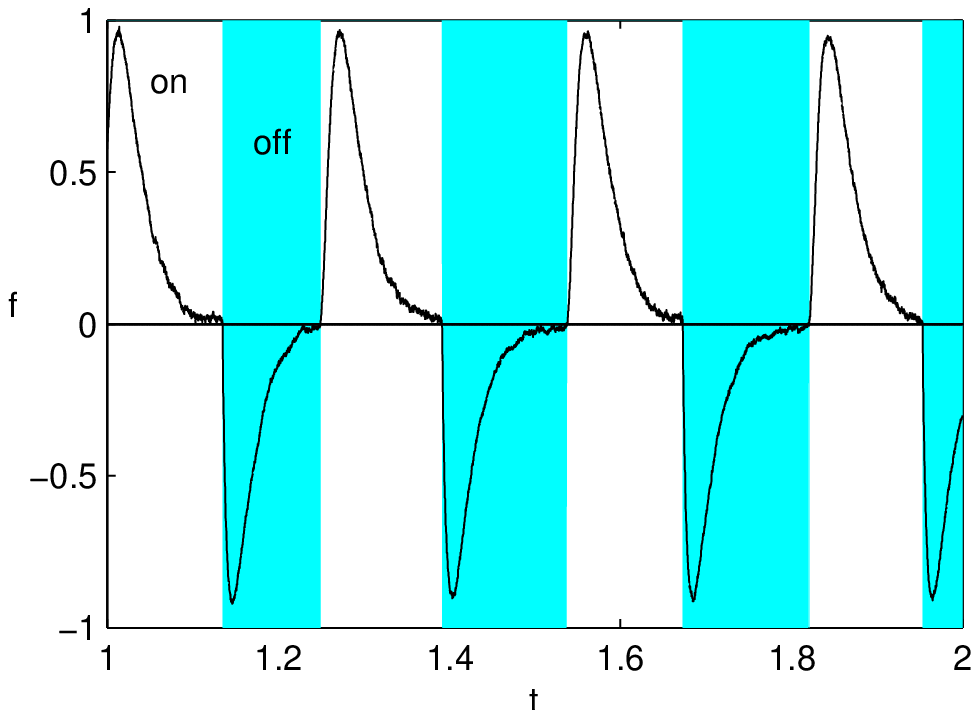}
\hspace{0.3cm}
\includegraphics[height=3cm]{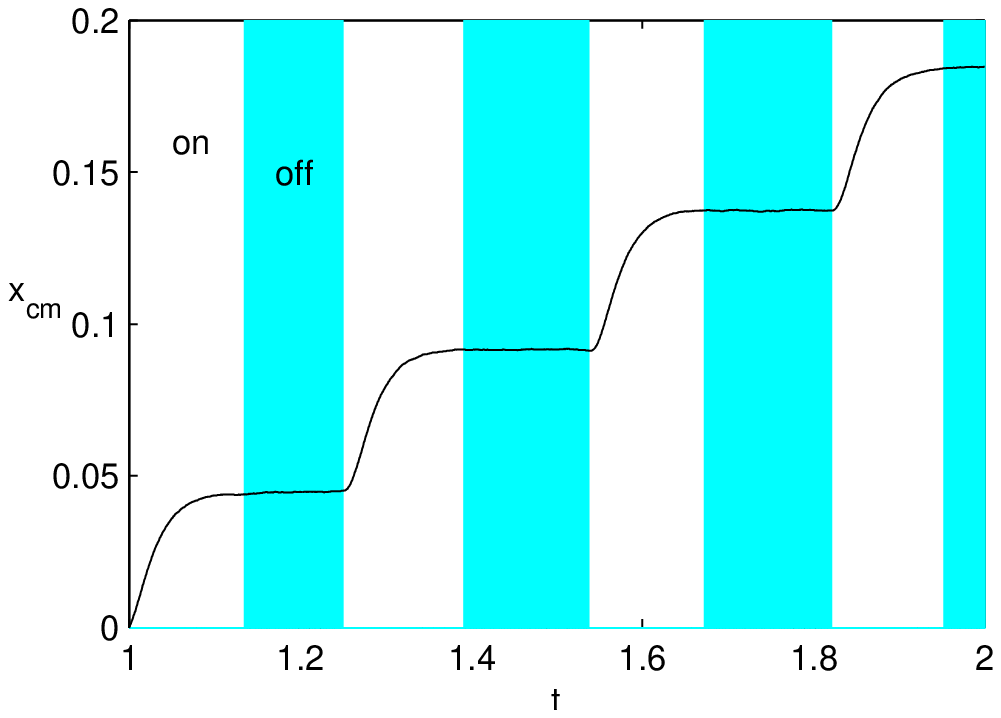}
 \]
\caption{$ N = 10^6$: Time dependence of the net force per particle $f(t)$
({\em left}) and the average position of the center of mass
$\langle x_{\rm cm}(t)\rangle$ ({\em right}). Colored background indicates
the off-potential time intervals. Both figures are for $V_0=5kT$, $a=1/3$.
(Units: $ L = 1 $, $ D = 1 $ and $ k T = 1 $.)} \label{fNevolF}
\end{figure}

For large $N$, numerical simulations reveal that the system gets
trapped near the equilibrium distribution of either the on or the
off potential, and switches are induced only by fluctuations of
the net force per particle $f(t)$. This behavior is shown in
Fig.~\ref{fNevolF}, where the evolution of $f(t)$ and $\langle
x_{\rm cm}(t)\rangle$ is depicted for $ N = 10^6 $. These
fluctuations decrease with $N$, then so does the frequency of
switches, yielding $\langle \dot x_{\rm cm} \rangle\to 0$ for
$N\to\infty$.
Therefore, the system almost reaches the equilibrium distribution
of each dynamics. These equilibrium distributions for the on and
off potential are, respectively, $\rho_{\rm st, on}(x)=
e^{-V(x)/kT}/Z$ and $\rho_{\rm st,off}(x)=1/L$, where $Z$ is a
normalization constant.

\begin{figure} [ht]
\[
\includegraphics[height=3cm]{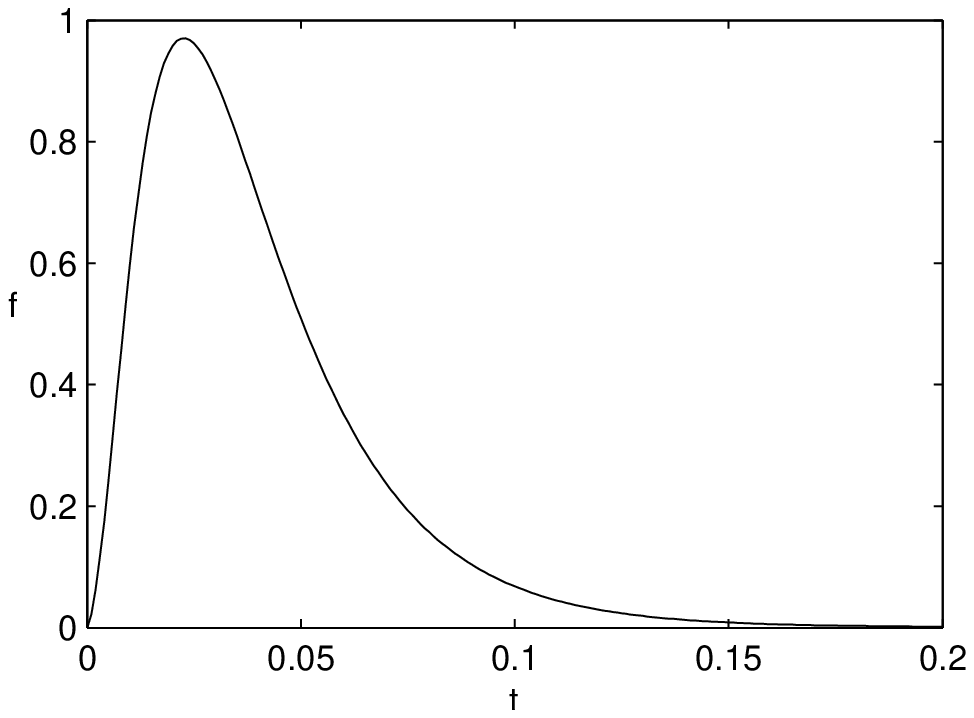}
\hspace{0.3cm}
\includegraphics[height=3cm]{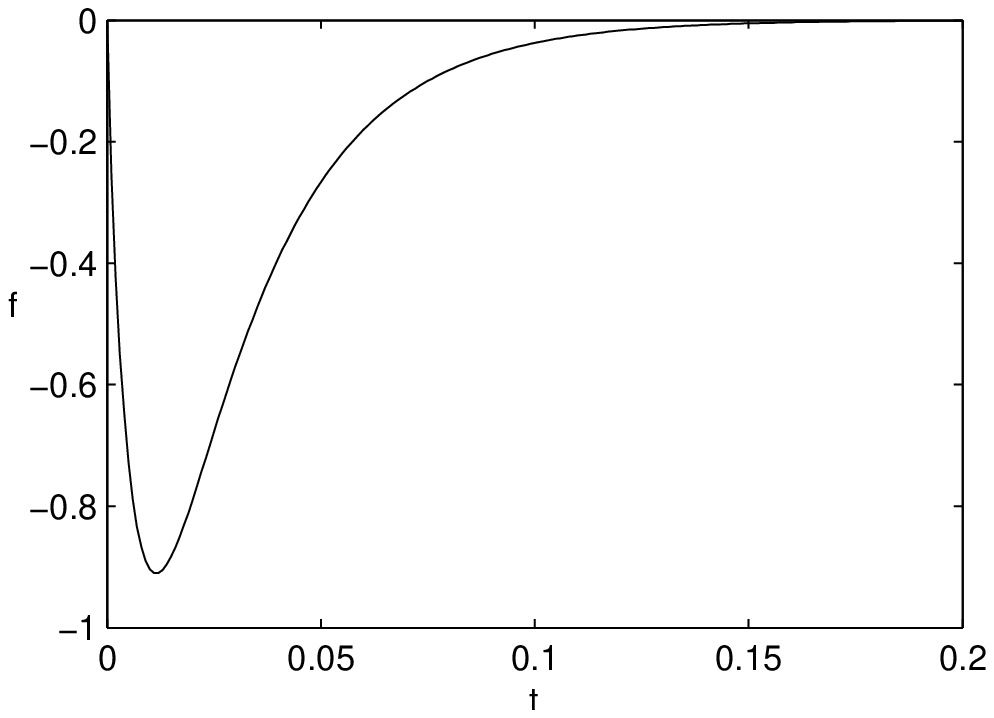}
\]
\caption{$ N = \infty $: Net force per particle $f^\infty(t)=\langle F(x)\rangle$ as a
function of time, for $V_0=5kT$, and $a=1/3$, and  for
the two cases: {\em Left}: $\alpha(t)=1$ (potential on) and
initial condition given by the equilibrium profile for a flat
potential, i.e., $\rho_{\rm st, off}(x)$; and {\em right}:
$\alpha(t)=0$ (potential off) and initial condition given by the
equilibrium profile for a ratchet potential, i.e.,  $\rho_{\rm st,
on}(x)$. (Units $ L = 1 $, $ D = 1 $, and $kT=1$.) }
\label{fFoffon}
\end{figure}

To gain some quantitative insight, let us consider the case
$N=\infty$ (no fluctuations). In this case, the distribution of
particles coincides with the solution of the following mean-field
Fokker-Planck equation:
\begin{equation}
\gamma \partial_t \rho (x,t) = \left[- \alpha(t)
\partial_x F(x) +kT\partial^2_{x}\right] \rho (x,t), \label{FP}
\end{equation}
where $\alpha(t)=\Theta\left( f^\infty(t)\right)$, $ f^\infty(t)
\equiv \langle F(x)\rangle $, and the average $\langle .\rangle$
is taken over $\rho(x,t)$.

We have studied the evolution of $\rho(x,t)$ in two situations:
{\em (i)} $\alpha(t)=1$ (potential on) with initial condition
given by $\rho_{\rm st,off}(x)=1/L$ and {\em (ii)} $\alpha(t)=0$
(potential off) with initial condition given by $\rho_{\rm
st,on}(x)$. We have calculated the average force $f^\infty(t)$ as
a function of time by a numerical integration of the Fokker-Planck
equation. The results are plotted in Fig.~\ref{fFoffon}. When the
potential is switched on, the particles in the negative force
interval (where the force is stronger due to the asymmetry of the
potential) arrive earlier at the minima. This leads to an initial
increase in the net force per particle, $ f^\infty(t) $. [A
similar interpretation can be given to Fig.~\ref{fFoffon}
(right).]

For large but finite $N$, $f(t)$ fluctuates around $f^\infty(t)$
(cfr. figures \ref{fNevolF} and \ref{fFoffon}), i.e., $ f(t) =
f^\infty(t) + fluctuations $. Due to these fluctuations, $f(t)$
crosses zero and switches are induced. In fact, for large $N$ the
controlled ratchet behaves as an excitable system: fluctuations
induce the switches and, after one of these switches, the system
has to perform a large excursion for the force to go back to a
small value, suitable for a new ``excitation''. This picture
allows us to calculate $\langle\dot x_{\rm cm}\rangle$ for large $N$.
For $N$ finite, the fluctuations of $f(t)$ are of order
$\sqrt{\langle F^2\rangle_{\rm eq}/N}$.
Therefore, a switch is possible whenever
\begin{equation}
|f^\infty(t)| \sim \sqrt{\frac{\langle F^2\rangle_{\rm eq}}{N}}.
\label{fluc}
\end{equation}
In our case, fluctuations are the same for the on and off
situations: $ \langle F^2 \rangle_{\rm eq} = V_0^2/[L^2a(1-a)]$.

On the other hand, $ f^\infty(t) $ departs rapidly from zero,
reaches a maximum and finally exhibits an exponential decay, both
for the on and off cases (Fig.~\ref{fFoffon}, left and right,
respectively). Therefore, in each case
\begin{equation}
f^\infty_{\rm on,off}(t) = C_{\rm on, off} e^{-\lambda_{\rm
on,off}(t-\tau_{\rm on, off})}, \label{efon}
\end{equation}
where $C_{\rm on,off}$ are constants, $\tau_{\rm on, off}$ are the
transient times of each dynamics, and $1/\lambda_{\rm on,off}$ are
the characteristic times of the corresponding exponential decays.
Combining Eqs. \eqref{fluc} and \eqref{efon}, one obtains the
following switching time:
\begin{equation}
t_{\rm on, off} \simeq c_{\rm on,off}+\frac{\ln N}{2\lambda_{\rm
on,off}}, \label{times}
\end{equation}
where  $c_{\rm on,off}$ are constants depending on the transient
regime of each dynamics.

Finally, the center of mass \emph{only} moves when the potential
is on, covering a distance $\Delta x_{\rm on}$, as shown in
Fig.~\ref{fNevolF} (right\textsc{}), (when the potential is off, the
evolution is purely diffusive and $ \Delta x_{\rm off} = 0 $.)
Thus, the average velocity of the particles is
\begin{equation}
\left\langle \dot x_{\rm cm} \right\rangle =
  \frac{\Delta x_{\rm on} }{ t_{\rm on} + t_{\rm off} } \simeq
  \frac{\Delta x_{\rm on} }{ b + d \ln N } . \label{final}
\end{equation}
where $ b = c_{\rm on} + c_{\rm off} $ and $ d = (\lambda_{\rm on}
+ \lambda_{\rm off}) / ( 2 \lambda_{\rm on} \lambda_{\rm off}) $.
Every parameter in Eq.~\eqref{final} can be obtained from the
numerical integration of the corresponding Fokker-Planck equation
for $ N = \infty $. We have performed such an integration for $
V_0 = 5 k T $ and $ a = 1/3 $, yielding $ \Delta x_{\rm on} =
0.047 L $, $ \lambda_{\rm on} = 40 D $, $ \lambda_{\rm off} =
39 D $ and $ b = -0.071 D^{-1} $.
Theory and numerical simulations show a good agreement for large
values of $N$ (see Fig.~\ref{J}). This good
agreement has also been verified for other values of $ V_0 $, $ a
$, and for other potential shapes, for example $ V(x) = V_0
[\sin(2\pi x/L) +
\sin(4\pi x/L)] $.

It is interesting to note that the advance $ \Delta x_{\rm on} $
is mainly covered during a transient time, $ \tau_{\rm on} $, that
is of the order of half the optimal period for periodic switching.
Longer switching periods, as we have in the present controlled
ratchet for large $ N $, lead to lower performance. On the other
hand,
 for small $ N $ the controlled ratchet manages to use the information
provided by $ f $ to do {\em adequate} faster switches and increase
performance.

Notice that the feedback control studied in this Letter is itself a
rectification mechanism. Hence, for $a \neq 1/2$, there are two 
sources of spatial asymmetry in our system: the feedback control and 
the shape of the ratchet potential. Feedback control induces
a net flux even for symmetric potentials, $a=1/2$. In this case,
$f^\infty(t)=0$ and one only has to consider the fluctuations in  
$f(t)$. Consequently, the binomial approximation works well even 
for large values of $N$. Introducing an asymmetry in the potential,
$a<1/2$,  favors the flow of particles for any $N$ and changes the 
large $N$ behavior of $\langle \dot x_{\rm cm}\rangle$ from 
$1/\sqrt{N}$ to $1/\ln N$ [cfr. Eqs.~\eqref{velbin} and \eqref{final}].

Summarizing, we have computed the current induced in an ensemble
of Brownian particles by a potential that is switched on and off
according to a simple feedback strategy maximizing the instant
velocity of the center of mass, and compared it with the current
induced by a periodic switching. The results show that, for small
$ N $, the current is better in the controlled case than in the
periodic switching, as expected. However, for large ensembles the
controlled ratchet performs worse than the periodic strategy.

There are a number of open problems prompted by our work. First,
finding the optimal protocol for $N>1$, which probably differs
from the one considered in this Letter. Secondly, we have focused
on maximizing the current, but it would be relevant to consider
also the maximization of the efficiency, specially to asses the
entropic value of the information about fluctuations and the
limitations imposed by thermodynamics to feedback control
\cite{lloyd}.

\acknowledgments
This work has been financially supported by grants
BFM2001-0291-C02-02 and BFM2003-02547/FISI from MCYT (Spain).
Authors also acknowledge support from UCM (Spain) through research
project PR1/03-11595, and from ESF Programme STOCHDYN.

\end{document}